\documentstyle[prl,aps,multicol,epsfig]{revtex}

\input epsf.tex

\newcommand{\be}{\begin{equation}}
\newcommand{\ee}{\end{equation}}
\newcommand{\ba}{\begin{eqnarray}}
\newcommand{\ea}{\end{eqnarray}}
\newcommand{\ban}{\begin{eqnarray*}}
\newcommand{\ean}{\end{eqnarray*}}

\newcommand{\sandwich}[3]{\mbox{$ \langle #1 | #2 | #3 \rangle $}}
\newcommand{\ket}[1]{\mbox{$ | #1 \rangle $}}

\newcommand{\demi}{\frac{1}{2}}

\newcommand{\one}{\leavevmode\hbox{\small1\normalsize\kern-.33em1}}

\newcommand{\moy}[1]{\langle #1 \rangle}

\begin{document}

\title{Entanglement and non-locality are different resources}
\author{Nicolas Brunner, Nicolas Gisin, Valerio Scarani}
\address{
Group of Applied Physics, University of Geneva, 20, rue de
l'Ecole-de-M\'edecine, CH-1211 Geneva 4, Switzerland}
\date{\today}
\maketitle \maketitle

\begin{abstract}
Bell's theorem states that, to simulate the correlations created
by measurement on pure entangled quantum states, shared randomness
is not enough: some "non-local" resources are required. It has
been demonstrated recently that all projective measurements on the
maximally entangled state of two qubits can be simulated with a
single use of a "non-local machine". We prove that a strictly
larger amount of this non-local resource is required for the
simulation of pure non-maximally entangled states of two qubits
$\ket{\psi(\alpha)}= \cos\alpha\ket{00}+\sin\alpha\ket{11}$ with
$0<\alpha\lesssim\frac{\pi}{7.8}$.
\end{abstract}

\begin{multicols}{2}

\section{Introduction}

There exists in nature a {\em channel} that allows to distribute
correlations between distant observers, such that (i) the
correlations are not already established at the source, and (ii)
the correlated random variables can be created in a configuration
of space-like separation, i.e. no normal signal can be the cause
of the correlations \cite{epr}. This intriguing phenomenon, often
called {\em quantum non-locality}, has been repeatedly observed,
and it is natural to look for a description of it. A convenient
description is already known: quantum mechanics (QM) describes the
channel as a pair of entangled particles. In the recent years,
there has been a growing interest in providing other descriptions
of this channel, mainly assuming a form of communication. Usually,
the interest in these description does not come from a rejection
of QM and the desire to replace it with something else: rather the
opposite, the goal is to quantify how powerful QM is by comparing
its achievements to those of other resources.

For instance, one may naturally ask how much information should be
sent from one party (Alice) to the other (Bob) in order to
reproduce the correlations that are obtained by performing
projective measurements on entangled pairs (to "simulate
entanglement"). The amount of communication is something that we
are able to quantify, thus the answer to this question provides a
measure of the non-locality of the channel. Bell's theorem implies
that some communication is required, but does not quantify this
amount. Several works \cite{canada,2bits} underwent the task of
estimating the amount of communication required to simulate the
maximally entangled state of two qubits (singlet). These partial
results were superseded in 2003, when Toner and Bacon \cite{toner}
proved that the singlet can be simulated exactly using local
variables plus {\em just one bit} of communication per pair. This
amount of communication is tight, in the absence of block-coding
--- which is indeed the way Nature does it: in an experiment, each
pair of entangled particles is "processed" independently of those
that preceded and those which will follow it.

More recently, another resource than communication has been
proposed as a tool to study non-locality: the {\em non-local
machine} (NLM) described by Popescu and Rohrlich \cite{pr},
sometimes called PR box --- actually, the first appearance of this
"machine" is Eq.~(1.11) of Ref.~\cite{tsi}. This hypothetical
machine was constructed to violate the Clauser-Horne-Shimony-Holt
(CHSH) inequality \cite{chsh} ${\cal B}\leq 2$ up to its algebraic
bound of ${\cal B}=4$ (while it is known that QM reaches up only
to ${\cal B}= 2\sqrt{2}$) without violating the no-signaling
constraint; and it would also provide a very powerful primitive
for information-theoretical tasks \cite{info,info2}. Cerf, Gisin,
Massar and Popescu \cite{machinesim} have shown that the singlet
can be simulated by local variables plus just a single use of the
NLM per pair: this is the analog of the Toner-Bacon result for
communication. While the NLM is by far a less familiar object than
bits of communication, in the context of simulation of
entanglement it has a very pleasant feature: it automatically
ensures that the no-signaling condition is respected. On the
contrary, bits of communication imply signaling: to reproduce
quantum correlations, as in the Toner-Bacon model, one must
cleverly mix different communication strategies in order to hide
the existence of communication. The idea itself of hidden
communication between quantum particles has several drawbacks
\cite{vs} and is hard to conciliate with the persistency of
correlations in experiments with moving devices \cite{andre}.

Thus, to date, the simulation of quantum non-locality has been
studied for two resources (communication and the NLM) and the
results are similar: the basic unit of the resource (one bit, or a
single use of the NLM) is sufficient for the simulation of the
singlet. Very few is known beyond the case of the singlet. Even
staying with just two qubits, the only known result is that two
bits of communication are enough to simulate all states
\cite{toner}, but this is not claimed to be tight. In this paper,
we study the analog problem using the NLM and demonstrate that, in
order to simulate the correlations of some pure non-maximally
entangled state of two qubits, a single use of the NLM is {\em
not} sufficient: an amount {\em strictly larger} of non-local
resources is needed than for the simulation of the maximally
entangled state. Curious as it may seem, this is not the first
example in the literature where entanglement and non-locality
don't behave monotonically with one another: Eberhard proved that
non-maximally entangled states require lower detection
efficiencies than maximally entangled ones, in order to close the
detection loophole \cite{eberhard}; Bell inequalities have been
found whose largest violation is given by a non-maximally
entangled state \cite{bineqs} and this has some consequences on
the communication cost as well \cite{pir}; it is also known that
some mixed entangled states admit a local variable model, even for
the most general measurements \cite{barr1}.

The present paper is structured as follows. As a necessary
introduction, we start by recalling the meaningful mathematical
tools for this investigation (Section \ref{sectools}). In Section
\ref{secmain} we demonstrate the main claim, by showing that there
exist a unique Bell-type inequality using {\em three} settings for
both Alice and Bob which is not violated by any strategy using the
NLM at most once, and which {\em is} violated by all the states of
the form $\ket{\psi(\alpha)}=
\cos\alpha\ket{00}+\sin\alpha\ket{11}$ for
$0\lesssim\alpha\lesssim\frac{2}{19}\pi$ (the sign of approximate
inequality means that these are numerical, not analytical
results). Thus, these states cannot be simulated by a single use
of the NLM. In the same Section, we show how this new inequality
can be violated by two uses of the NLM or by one bit of
communication, and comment on these features. In Section
\ref{secmain2} we consider extension to more settings on Alice's
and/or on Bob's side. The case of {\em four} settings for Alice
and three settings for Bob allows us to extend the result to the
range $0<\alpha\lesssim\frac{\pi}{7.8}$ (in particular, we prove
that at least two uses of the NLM are required to simulate pure
states arbitrary close to the product state $\alpha=0$). The
(admittedly incomplete) survey of other cases did not provide
further improvements. Section \ref{seccon} is a conclusion.

\section{Tools: polytopes and the no-signaling condition}
\label{sectools}

\begin{figure}
\epsfxsize=8cm \epsfbox{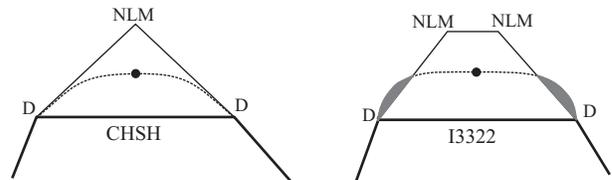} \caption{Representation of
the regions in probability space. The thick line represents the
polytope of shared randomness; its vertices are deterministic
strategies (D). Above it lies the new polytope obtained with a
single use of the NLM. The curved line encloses the points
achievable with measurements on quantum states. The black dot
represents the measurement on the singlet that gives the maximal
violation of the inequality corresponding to the facet (CHSH or
$I_{3322}$). The existence of the grey regions above the
$I_{3322}$-facet is the result of Section
\ref{secmain}.}\label{figpoly}
\end{figure}

Instead of tackling the issue of simulating all possible
measurements done on an entangled state, we consider a restricted
protocol, as typical in Bell's inequalities. Obviously, if this
restricted protocol cannot be simulated, {\em a fortiori} it will
be impossible to simulate all the correlations. We allow then each
of the two physicists, called Alice and Bob, to choose between a
finite set of possible measurements $\{A_i\}_{i=1...m_A}$,
$\{B_j\}_{j=1...m_B}$. As a result of each measurement on a pair,
they get an outcome noted $r_A$, $r_B$. We focus here on
dichotomic observables (like von Neumann measurements on qubits),
with the convention $r_{A,B}\in\{0,1\}$. An "experiment" is fully
characterized by the family of probabilities
$P(r_A,r_B|A_i,B_j)\equiv P_{ij}(r_A,r_B)$. There are $d=4m_Am_B$
such probabilities, so each experiment can be seen as a point in a
region of a $d$-dimensional space, bounded by the conditions that
probabilities must be positive and sum up to one. By imposing
restrictions on the possible distributions, the region of possible
experiment shrinks, thus adding non-trivial boundaries
\cite{tsi,pito,barrett}. For instance, one may require that the
probability distribution must be built without communication, only
with shared randomness. In this case, the bounded region actually
forms a {\em polytope}, that is a convex set bounded by
hyperplanes ("facets") which are Bell's inequalities in the usual
sense. The vertices of this polytope are {\em deterministic
strategies}, that is, probability distributions obtained by
setting $r_{A,B}$ always at 0 or always at 1 for each setting. The
vertices are thus easily listed, but to find the facets given the
vertices is a computationally hard task. The probability
distributions obtained with a single use of the NLM also form a
polytope, obviously larger than the one of shared randomness ---
actually, the vertices associated to deterministic strategies
remain vertices of this new polytope, but more vertices are added.
Finally, the probability distributions that can be obtained from
measurements on quantum states form a convex set which is not a
polytope \cite{note3}. A sketch of this structure is given in
Fig.~\ref{figpoly}, which will be commented in more detail in the
following.

Since our goal is to simulate QM, we impose from the beginning the
constraints of {\em no-signaling}; that is, we focus only on those
probability distributions which fulfill \ba
\sum_{r_A}P_{ij}(r_A,r_B)&=&P_j(r_B)\;\;\mbox{for all $i$} \ea and
a similar condition for the marginal of A. Under no-signaling, the
full probability distribution is entirely characterized by
$d_{ns}=m_Am_B+m_A+m_B$ probabilities, which we choose
conventionally to be the $P_i(r_A=0)$, $P_j(r_B=0)$ and
$P_{ij}(r_A=r_B=0)$ as in Ref. \cite{collins}.

\section{Main result}
\label{secmain}

\subsection{Basic notations}

Let's focus more specifically on the first case of interest for
this paper, $m_A=m_B=3$. The no-signaling probability space is
15-dimensional. All the facets of the deterministic polytope are
known \cite{collins}: up to relabelling of the settings and/or of
the outcomes, they are equivalent either to the usual two-settings
CHSH inequality, or to the truly three-settings inequality
$I_{3322}$ that reads \ba I_{3322}\,=\,
\begin{array}{c|ccc} & -1&0&0\\\hline -2&1&1&1\\ -1&1&1&-1\\ 0&1&-1&0\\ \end{array}&\leq &0\,.
\label{i3322}\ea Here the notation represents the coefficients
that are put in front of the probabilities, according to \ba
\begin{array}{c|c} & P_i(r_A=0)\\\hline P_j(r_B=0)&
P_{ij}(r_A=r_B=0)\,.
\end{array}\ea The maximal violation allowed by QM is obtained for the singlet
and is $\moy{I_{3322}}=\frac{1}{4}$. To become familiar with the
notations, the deterministic strategy in which Alice outcomes
$r_A=0$ for $A_0,A_1$ and $r_A=1$ for $A_2$, and Bob outcomes
always $r_B=0$, corresponds to the probability point \ba
[0_d\,0_d\,1_d; 0_d\,0_d\,0_d]&=&
\begin{array}{c|ccc} & 1&1&0\\\hline 1&1&1&0\\ 1&1&1&0\\ 1&1&1&0\\
\end{array}\,.\ea To see the result of $I_{3322}$ on this
strategy, one has simply to multiply the arrays term-by-term: here
we find $I_{3322}=0$. There are obviously $2^6=64$ deterministic
strategies; among these, 20 saturate the inequality $I_{3322}=0$
(i.e., they lie on the facet) while the others give $I_{3322}<0$.
To verify that this is indeed a facet, it is enough to show that
the rank of the matrix containing the 20 points that saturate the
inequality is $d_{ns}-1=14$, so that the condition $I_{3322}=0$
really defines a hyperplane \cite{masanes}.

\subsection{The polytope of a single use of the non-local machine
(NLM)}

\begin{figure}
\epsfxsize=7cm \epsfbox{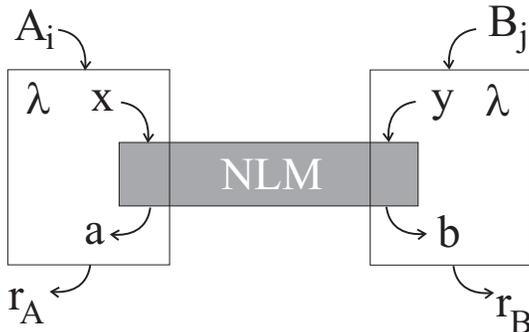} \caption{Schematics of a
strategy allowing a single use of the NLM between Alice and Bob.
See text for details.}\label{figmach}
\end{figure}

The NLM is defined as a two-input and two-output channel. Alice
inputs $x$ and gets the outcome $a$, Bob inputs $y$ and gets the
outcome $b$; all these numbers take the values 0 or 1, the
marginal distribution on each side is completely random,
$P(a=0)=P(b=0)=\frac{1}{2}$, but the outcomes are correlated as
\ba a+b&=&xy\,.\ea Explicitly, if either $x=0$ or $y=0$, then
$(a,b)=(0,0)$ or $(1,1)$ with equal probability; if $x=y=1$, then
$(a,b)=(0,1)$ or $(1,0)$ with equal probability.

The most general strategy allowing a single use of the NLM is
sketched in Fig.~\ref{figmach}. Alice and Bob share some random
variable $\lambda$. Alice inputs in the machine the bit
$x=x(A_i,\lambda)$; the machine gives the output $a$, and Alice
outputs $r_A=r_A(A_i,\lambda,a)$. The extremal strategies are such
that a given $x$ is associated to each $A_i$, and the outcome is
either $r_A=a$ or $r_A=1-a$. Similarly for Bob. Note that it is
also possible that for a given pair $(A_i,B_j)$ Alice uses the
machine while Bob outputs a deterministic bit; in this case, we
can suppose that Bob inputs $y=0$ in the machine but does not use
the output $b$. We shall come back below to the listing of
extremal strategies. Let's now see how the polytope of possible
probability distributions is enlarged by allowing a single use of
the NLM.

By construction, with a single use of the machine one can violate
the CHSH inequality more than is possible in QM. In the polytope
picture, one new point appears above any face corresponding to
CHSH: the facets of the enlarged polytope should now pass through
this points. They must pass through the deterministic points as
well, because these points are still extremal. All this is
sketched in Fig.~\ref{figpoly}. We won't study this example
longer, however, since there is no hope of finding something
interesting above a CHSH-like facet: all the probability
distributions involving two settings which are no-signaling (in
particular all distributions arising from measurements on a
quantum state) can be simulated by a single use of the NLM
\cite{barrett}.

So let's consider the facet defined by $I_{3322}=0$. Again, a
single use of the NLM allows to violate $I_{3322}$, which is
expected since the NLM can in particular simulate the singlet. For
instance, consider the strategy in which Alice inputs $x=0$ in the
NLM for $A_0$ and $A_1$, and $x=1$ for $A_2$, while Bob inputs
$y=0$ for $A_0$ and $B_2$, and $y=1$ for $B_1$; in each case, the
outputs are $r_A=a$, $r_B=b$. This strategy gives the probability
point \ba [0_m\,0_m\,1_m; 0_m\,1_m\,0_m]&=&
\frac{1}{2}\times\,\begin{array}{c|ccc} & 1&1&1\\\hline 1&1&1&1\\ 1&1&1&0\\ 1&1&1&1\\
\end{array} \label{stratm}\ea yielding $I_{3322}=\frac{1}{2}$: the machine can violate $I_{3322}$
more than QM. However here, we are going to show what is
graphically represented in Fig.~\ref{figpoly}: for non-maximally
entangled states, some points achievable with quantum states lie
outside the enlarged polytope. In other words, the facets of this
polytope define generalized Bell's inequalities that can still be
violated by QM.

To find the facets of the polytope allowing a single use of the
machine, one must first list all the vertices (extremal
strategies). This can be done systematically on a computer, once
having noticed that for each setting, Alice and Bob have six
choices: deterministically output 0 or 1 (noted $0_d$, $1_d$),
input $0$ or $1$ in the machine and keep the output of the machine
(noted $0_m$, $1_m$), input $0$ or $1$ in the machine and flip the
output of the machine (noted $0_f$, $1_f$) \cite{note2}. This
listing gives {\em a priori} $6^6=46656$ strategies, although many
of them are equal \cite{note0} and only 3088 different strategies
are left after inspection. Note that some of these are not even
extremal points of the polytope: for instance, the strategy
$[0_m,0_m,0_m;0_m,0_m,0_m]$, in which both Alice and Bob input 0
in the NLM for all settings, yields the same probability point as
the equiprobable mixture of the deterministic strategies
$[0_d,0_d,0_d;0_d,0_d,0_d]$ and $[1_d,1_d,1_d;1_d,1_d,1_d]$.
Certainly not equivalent to a mixture of deterministic strategies,
however, are those strategies like (\ref{stratm}) which violate
$I_{3322}$: upon counting, there are 28 of these, all giving the
same violation $I_{3322}=\frac{1}{2}$. To find the facets of the
new polytope, we use available computer programs \cite{matlab}. We
find some trivial facets, plus a single non-trivial one
\cite{twofaces} which reads \ba M_{3322}\,=\,
\begin{array}{c|ccc} & -2&0&0\\\hline -2&1&1&1\\ -1&1&1&-1\\ 0&1&-1&0\\ \end{array}&\leq &0\,.
\label{m3322}\ea This new inequality is extremely similar to
$I_{3322}$, eq.~(\ref{i3322}): only the coefficient of
$P_{0}(r_A=0)$ is now $-2$ instead of $-1$. The origin of this
difference can be appreciated, at least to some extent: one of the
biggest difficulties found in adapting the Toner-Bacon model
\cite{toner} to non-maximally entangled states lies in the need of
simulating not only the correlations, but also the non-trivial
marginal distributions. It is thus a good idea for our purpose, to
add penalties on the marginal distributions.

Summarizing, we have found a tight inequality $M_{3322}\leq 0$
which is satisfied by all the 3088 extremal points of the polytope
of probabilities achievable by shared randomness plus a single use
of the NLM. Now we move on to show that QM violates this
inequality.

\subsection{Violation of the inequality with pure non-maximally entangled states}

\begin{figure}
\epsfxsize=8cm \epsfbox{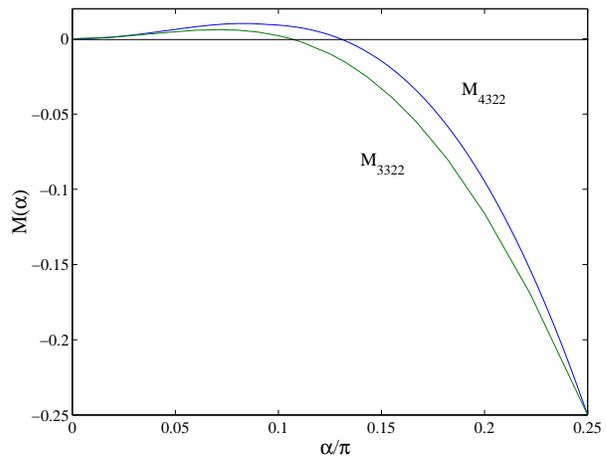} \caption{Value of the
quantum-mechanical expectation $M(\alpha)$ as a function of
$\alpha$ for the optimized settings of Alice and Bob, for the
inequalities $M_{3322}$, Eq.~(\ref{m3322}), and $M_{4322}$,
Eq.~(\ref{m4322}). The region where $M(\alpha)>0$ corresponds to
the grey regions in Fig.~\ref{figpoly}: the corresponding quantum
states cannot be simulated by a single use of the
NLM.}\label{figopti}
\end{figure}

We consider states of the form \ba \ket{\psi(\alpha)}&=&
\cos\alpha\ket{00}+\sin\alpha\ket{11}\ea with
$\cos\alpha\geq\sin\alpha\geq 0$. Up to local operations, this is
the most general pure state of two qubits (Schmidt decomposition).
We form the Bell operator ${\cal M}$ as usual, by replacing the
probabilities in (\ref{m3322}) by the corresponding
one-dimensional projectors, like \ba P_0(r_A=0) &\longrightarrow &
\demi\left(\one+\vec{a}_0\cdot\vec{\sigma}\right) \label{proj}\ea
For each $\alpha$, we have to find the settings which maximize
$M(\alpha)=\sandwich{\psi(\alpha)}{{\cal M}}{\psi(\alpha)}$. We
have not found a closed analytical formula, but it is easy for a
computer to optimize over twelve real parameters. The result is
shown in Fig.~\ref{figopti}: one finds $M(\alpha)>0$ for
$0<\alpha\lesssim\frac{2}{19}\pi$, with a maximal violation
$M(\bar{\alpha})\approx 0.0061$ at $\bar{\alpha}\approx
0.0712\,\pi$. It appears that all the optimal settings are of the
form $\hat{a}_i = \cos\theta_{A}^{i}\hat{z}+\sin\theta_{A}^{i}
\hat{x}$, $\hat{b}_j = \cos\theta_{B}^{j}\hat{z}+\sin
\theta_{B}^{j}\hat{x}$.

The curve of Fig.~\ref{figopti} is the exact version of the
pictorial argument of Fig.~\ref{figpoly}. Note in particular the
following features: (i) As expected, there is no violation for the
singlet ($\alpha=\frac{\pi}{4}$), because this state can be
simulated with the NLM; even more, $M(\frac{\pi}{4})=-\frac{1}{4}$
which is the difference between $\moy{I_{3322}}=\frac{1}{4}$ on
the singlet and the maximal value $I_{3322}=\frac{1}{2}$
achievable with the NLM. The picture of Fig.~\ref{figpoly} yields
in this case even a quantitative prediction. (ii) As mentioned
\cite{note3}, it is not obvious that the set of probabilities
obtained from quantum measurements on $\ket{\psi(\alpha)}$ is
convex; so at this point it is not proved that states arbitrarily
close to the product state $\ket{00}$ can't be simulated by a
single use of the NLM --- the proof will be provided in Section
\ref{secmain2}.

\subsection{Violation of the inequality with two NLMs}
\label{ss1}

As we mentioned in the introduction, two bits of communications
are a sufficient resource to simulate any state of two qubits. The
analog simulation using twice the NLM is still missing, and may
even not exist. While waiting for more clarification, we have
found a way of violating inequality (\ref{m3322}) by using the NLM
twice. The settings are coded as $A\to(x',x'')$, $B\to(y',y'')$
according to: $A_0\to(0,0)$, $A_1\to(0,1)$, $A_2\to(1,0)$;
$B_0\to(0,0)$, $B_1\to(1,0)$ and $B_2\to(0,1)$. Then, $x'$ and
$y'$ are used as inputs in the first use of the machine, whose
outcomes are denoted $a'$ and $b'$; $x''$ and $y''$ are used as
inputs in the second use of the machine, whose outcomes are
denoted $a''$ and $b''$. Finally, Alice outputs $r_A=a'+a''$, Bob
outputs $r_B=b'+b''$ (sum modulo 2). This strategy gives the
probability point \ba [\mbox{Two uses of the NLM,...}]&=&
\frac{1}{2}\times\,\begin{array}{c|ccc} & 1&1&1\\\hline 1&1&1&1\\ 1&1&1&0\\ 1&1&0&1\\
\end{array}\ea yielding $I_{3322}=1$ and $M_{3322}=\demi$.

We can go a step further. Consider a mixture of two strategies:
with probability $p$, the strategy just described which uses two
NLMs; with probability $1-p$, the deterministic strategy in which
Alice and Bob output always $r_A=r_B=1$ for all settings (all
entries of the table are zeros), and which obviously does not
require any use of the NLM. Such a mixed strategy yields a
violation $M_{3322}=\frac{p}{2}$ of the inequality. Now, if
$p<\demi$, this strategy uses less than one NLM {\em on average}
\cite{pironio}. This is not a contradiction with our main result:
at least two NLMs must be available to simulate non-maximally
entangled states, albeit possibly this resource is not used for
all items.

\subsection{Violation of the inequality with one bit of communication}
\label{ss2}

It is natural to ask whether inequality (\ref{m3322}) provides
also a Bell inequality for one bit of communication. Bacon and
Toner \cite{bacon} had studied such inequalities for three
settings, but they restricted to correlation inequalities, whereas
inequality (\ref{m3322}) is a probability inequality. The problem
is complex because, as we mentioned, pure strategies with one bit
of communication imply signaling; we must find a mixture of such
strategies which is no-signaling and which violates our
inequality. It turns out that such mixtures do exist, so that our
inequality is {\em not} an inequality for one bit of
communication. In other words, the polytope of the probability
distributions obtained with one bit of communication plus the
no-signaling constraint is larger than the one associated to a
single use of the NLM.

As an explicit example \cite{pironio}, it can be verified that the
no-signaling strategy \ba [\mbox{One bit, no-signaling}]&=&
\frac{1}{5}\times\,\begin{array}{c|ccc} & 1&1&1\\\hline 1&1&1&1\\ 1&1&1&0\\ 1&1&0&1\\
\end{array}\ea yields the violation $M_{3322}=\frac{1}{5}$ and can be
obtained as the equiprobable mixture of the following five one-bit
strategies: \ba
%[r_{A_0},r_{A_1},r_{A_2};r_{B_0},r_{B_1},r_{B_2}|c(A_0),c(A_1),c(A_2)]\,=\nonumber\\
[\{r_{A_i}\};\{r_{B_j}\}|c_0,c_1,c_2]&=&\left\{\begin{array}{l} \,[1,1,0;c,1,c|1,1,0]\\
\,[1,0,1;c,c,1|1,0,1]\\
\,[0,1,1; c,c,c|0,1,1]\\
\,[1,1,1; 1,1,c|1,0,1]\\
\,[1,1,1; 1,c,1|1,1,0]
\end{array}\right.\,. \ea In these notations, $c_i$ is the
value of the bit that Alice sends to Bob when she has used the
setting $A_i$; $r_{B_j}=c$ means that, upon choosing the setting
$B_j$, Bob outputs the value of the bit received from Alice.

The fact that the inequality (\ref{m3322}) can be violated by one
bit of communication shows that a single use of the NLM does not
correspond to a single bit of communication plus no-signaling: the
NLM is a resource strictly weaker than communication, as argued in
Ref.~\cite{machinesim}, and grasps finer details of the structure
of quantum non-locality. The question whether one bit of
communication is sufficient to simulate non-maximally entangled
states is obviously still open.

\section{Extensions to more settings}
\label{secmain2}

In the previous Section, we have provided a complete study of the
case $m_A=m_B=3$: there cannot be any inequality other than
(\ref{m3322}) which has the desired properties. In this Section,
we explore other cases, starting from the next easiest, namely
$m_A=4$ and $m_B=3$.

\subsection{The case $m_A=4$ and $m_B=3$}

In the case $m_A=4$ and $m_B=3$, the no-signaling probability
space is 19-dimensional. All the facets of the deterministic
polytope have been listed in Appendix A of Ref.~\cite{collins}:
one finds of course CHSH, $I_{3322}$, plus three new inequalities.
The one which turns out to be of interest is (A2) of that
reference \cite{note1}; using the properties of no-signaling
distributions, and providing Alice instead of Bob with the four
settings, this inequality can be re-written in the form \ba
I^{(2)}_{4322}&=&
\begin{array}{c|cccc}
& -1&0&0&0\\\hline -2&1&1&1&0\\ -1&1&0&-1&1\\ 0&1&-1&0&-1\\
\end{array}\,\leq 0\,. \label{i4322}\ea
The polytope of a single use of the NLM can be found as in Section
\ref{secmain}. After listing, one finds that the $6^7$ possible
extremal strategies produce 17272 different points, 63 of which
violate (\ref{i4322}) by $I^{(2)}_{4322}=\demi$. By numerical
inspection \cite{note4}, we found that these points define a
single non-trivial new facet \ba M_{4322}&=&
\begin{array}{c|cccc}
& -2&0&0&0\\\hline -2&1&1&1&0\\ -1&1&0&-1&1\\ 0&1&-1&0&-1\\
\end{array}\,\leq 0\,. \label{m4322}\ea Quite similarly to the
case $m_A=m_B=3$, the difference between the original inequality
and the new one is just a larger penalty on one marginal. Also
similar is the fact that (\ref{m4322}) can be violated by
strategies which use twice the NLM or one bit of communication, as
can be easily verified. In fact, simply by taking the
corresponding strategies of Section \ref{secmain} and adding the
condition $A_3=A_1$, we can produce the probability point \ba
[\mbox{two NLMs, one bit}]&=&\lambda\times\,\begin{array}{c|cccc} & 1&1&1&1\\
\hline 1&1&1&1&1\\ 1&1&1&0&1\\ 1&1&0&1&0\\
\end{array}\ea which gives $M_{4322}=\lambda$, with $\lambda = \demi$ in the case of two
NLMs and $\lambda = \frac{1}{5}$ in the case of one bit of
communication.

The interest of $M_{4322}$ comes from the quantum violation, which
(i) is larger than the violation of $M_{3322}$, thus allowing to
extend the range of $\alpha$ for which one NLM is not enough, and
(ii) is obtained for a family of settings which can be easily
parametrized (see Appendix \ref{appa}). Specifically, one finds
$M(\alpha)>0$ for $0<\alpha\lesssim\frac{\pi}{7.8}$, with a
maximal violation $M(\bar{\alpha})\approx 0.0102$ at
$\bar{\alpha}\approx \frac{\pi}{12}$ (Fig. \ref{figopti}). For
small values of $\alpha$, moreover, one can prove \ba
M(\alpha)\geq \frac{1}{4}\,\alpha^2+O(\alpha^4)\,.\label{bound}\ea
Thus $M(\alpha)>0$ as soon as $\alpha>0$: the simulation of pure
states with arbitrarily weak entanglement requires more than one
NLM --- again, as we noticed at the end of paragraph \ref{ss1}, it
may be the case that correlations can be reproduced by using this
resource only on a subset of the particles.

\subsection{Other inequalities}

Beyond the $3322$ and $4322$ cases, the facets of the
deterministic polytope have not been listed exhaustively, but
several examples of facets are available \cite{collins,collins2}.
On these, we searched for possible extensions of our results by
increasing the penalties in some marginals.

Starting from the inequality $I_{4422}\leq 0$ given in Ref.
\cite{collins}, the corresponding inequality $M_{4422}\leq 0$ is
obtained exactly as above, just replacing $-1$ with $-2$ as the
coefficient of $P_{0}(r_A=0)$.The result is similar: $M_{4422}\leq
0$ indeed holds for all strategies allowing a single use of the
NML, and QM violates it. If all the four settings are used, the
range of values of $\alpha$ in which we found a violation is
however {\em smaller} than for $M_{3322}$, only up to
$\sim\frac{\pi}{13}$. Note that $I_{4422}$ is less violated than
$I_{3322}$ by the singlet \cite{collins}, while the violation
achievable by the NLM is $\demi$ for both; by looking at
Fig.~\ref{figpoly}, it becomes intuitive that the range of
violation should decrease. Interestingly, one can recover the
(better) result of Fig.~\ref{figopti} by setting $a_3$ and $b_0$
to the value $1_d$, thus reducing $M_{4422}$ to $M_{3322}$. This
assignment reads $P_3(r_A=0),P_0(r_B=0)\rightarrow 0$ and is thus
{\em not} of the form (\ref{proj}): it describes a degenerate
measurement.

Other 4422 deterministic facets, as well as some 5522 and 6622
ones, did not appear to be worth a closer study after our survey.
We have considered neither inequalities with larger number of
outcomes, nor multi-partite scenarios.

\section{Conclusion and perspectives}
\label{seccon}

In conclusion, we have shown that the simulation of non-maximally
entangled states of qubits requires a strictly larger amount of
resources (use of the non-local machine) than the simulation of
the singlet.

We have completely solved the problem in the case of three
settings and two outcomes for both Alice and Bob, and found an
extension in the case where Alice chooses among four settings. At
present thus, we know that the singlet can be simulated by a
single use of the NLM, while the simulation of states with
$0<\alpha\lesssim \frac{\pi}{7.8}$ requires that more than one NLM
is available (even though possibly this resource is used only
seldom). It will be of great interest to fill the gap, and to see
whether a similar result holds when the resource used to simulate
correlation are bits of communications instead of the NLM.

From a very fundamental point of view, we have discovered a new
surprising feature of the quantum world, which shows once more how
far this world lies from our intuition. But a precise
understanding of the incommensurability between entanglement and
non-locality would be of interest for applications as well. For
instance, it would allow to study whether in a given quantum
information protocol (cryptography, teleportation, an
algorithm...) it is better to look for the largest amount of
entanglement or for the largest amount of non-locality.

We acknowledge financial support from the EU Project RESQ and from
the Swiss NCCR "Quantum photonics".

\appendix

\section{Optimal settings for $M_{4322}$}
\label{appa}

To compute the maximal violation of $M_{4322}$, Eq.~(\ref{m4322}),
on qubit states, one must perform an optimization over 14
parameters. This we first performed numerically; by looking at the
result however, an analytical form for the settings has been
guessed. We give the settings by indicating the azymutal and polar
angle of the vector in the Bloch sphere
$\hat{n}\equiv(\theta,\varphi)$.

For $0\leq\alpha \lesssim \frac{\pi}{10.6}$, the optimal settings
lie in the $(x,z)$ plane and only two parameters depend on $\alpha$; specifically
\ban A_0&=&(\pi,0)\,=\,-\hat{z}\\
A_1&=&(\theta_A,\pi) \\
A_2&=&(\frac{\pi}{2},\pi)\,=\,-\hat{x}\\
A_3&=&(\theta_A,0)\\
B_0&=&(\theta_B,\pi)\\
B_1&=&(\theta_B,0)\\
B_2&=&(\pi,0)\,=\,-\hat{z}\,.\ean This gives \ban
M(\alpha)&=&\demi\,\Big[-3+\cos 2\alpha+\cos\theta_A\\
&&-\cos\theta_B(1+\cos 2\alpha)+ \cos\theta_A\cos\theta_B\\
&&+\sin\theta_B(1+\sin\theta_A)\sin 2\alpha\Big]\,.\ean We have
not been able to find a closed formula for $\theta_{A,B}$ as a
function of $\alpha$.

For $\frac{\pi}{10.6}\lesssim \alpha\leq \frac{\pi}{4}$, the
optimal settings don't lie in the $(x,z)$ plane any longer, and
only one parameter depends on $\alpha$; specifically
\ban A_0&=&(\pi,0)\,=\,-\hat{z}\\
A_1&=&(\frac{\pi}{2},\frac{5\pi}{6}) \\
A_2&=&(\frac{\pi}{2},\frac{\pi}{2})\,=\,+\hat{y}\\
A_3&=&(\frac{\pi}{2},\frac{7\pi}{6})\\
B_0&=&(\theta_B,\frac{4\pi}{3})\\
B_1&=&(\theta_B,\frac{2\pi}{3})\\
B_2&=&(\theta_B,0)\,.\ean This gives \ban
M(\alpha)&=&\frac{1}{4}\,\Big[-7+\cos 2\alpha-3\cos\theta_B(1+\cos
2\alpha)\\&&+3\sqrt{3}\sin\theta_B\sin 2\alpha\Big]\ean which can
easily be maximized to find \ban
\theta_B(\alpha)&=&\pi-\mbox{Arctan}\left(\frac{\sqrt{3}\,\sin
2\alpha}{1+\cos 2\alpha}\right)\,. \ean Even though these are not
the best settings for small values of $\alpha$, we can study the
limit $\alpha\rightarrow 0$ and we find $M(\alpha)=
\frac{1}{4}\,\alpha^2+O(\alpha^4)$ which means a violation of the
inequality for arbitrary small values of $\alpha$. Thus, we prove
analytically that at least this value can be reached, as written
in the main text, Eq.~(\ref{bound}).

\end{multicols}


\begin{thebibliography}{10}
\bibitem{note2} This takes automatically into account the possibility that Alice and Bob use the alternative
version of the NLM defined by $a+b=xy+1$.
\bibitem{matlab} We used the function \verb"convhulln" of \textsc{Matlab} on the
28 points which yield $I_{3322}=\frac{1}{2}$ alone, or on these
points plus some of the 20 deterministic points which yield
$I_{3322}=0$. The result is the same.
\bibitem{twofaces} Actually, the program outputs two facets:
inequality (\ref{m3322}), and a similar one where the first line
(2,0,0) is replaced by (1,1,0). If one writes down the inequality
explicitly, it becomes evident that one is transformed into the
other by exchanging $A_0$ and $A_1$ and by flipping the bit
$r_{B_2}$.
\bibitem{note1} Actually, we did not perform a systematic search
because the polytope becomes rather big. Based on the result of
Section \ref{secmain}, we have taken inequalities (A1)-(A3) of
\cite{collins}, verified that all of them can be violated by a
single use of the NLM, then modified the marginals. It turns out
that, even by modifying the marginals by 1, the new inequalities
derived from (A1) and (A3) are not violated by QM.
\bibitem{note4} Finding a polytope given 63 vertices is too hard a task
for the function \verb"convhulln" of \textsc{Matlab}. Therefore,
we have run this function more than one million of times, each
time randomly selecting 30 points out of the 63. When the selected
points defined a non-trivial facet, it was invariably
(\ref{m4322}). Thus we are confident that this is the only
interesting inequality for one use of the NLM above the
$I^{(2)}_{4322}$ deterministic facet.
\end{thebibliography}
\end{document}